\documentclass[final,5p,times,twocolumn]{elsarticle}

\usepackage{amssymb}
\usepackage{amsmath}
\usepackage{bm}
\usepackage{bbm}

\journal{Physics Letters B}

\begin{document}

\begin{frontmatter}



\title{Convergence of the $ppp$ correlation function within the hyperspherical adiabatic basis}

 \author[1]{E. Garrido}
 \author[2]{A. Kievsky}
 \author[3]{R. Del Grande}
 \author[4]{L. \v{S}erk\v{s}nyt\.{e}}
 \author[2]{M. Viviani}
 \author[2,5]{L.E. Marcucci}
 \affiliation[1]{organization={Instituto de Estructura de la Materia, CSIC},
             addressline={Serrano 123},
             city={Madrid},
             postcode={E-28006},
             country={Spain}}

 \affiliation[2]{organization={Istituto Nazionale di Fisica Nucleare},
             addressline={Largo Pontecorvo 3},
             city={Pisa},
             postcode={56127},
             country={Italy}}

 \affiliation[3]{organization={Faculty of Nuclear Sciences and Physical Engineering, Czech Technical University in Prague},
             addressline={Břehová 7},
             city={Prague},
             postcode={11519},
             country={Czech Republic}}
             
 \affiliation[4]{organization={Experimental Physics Department, CERN},
             city={Geneva 23},
             postcode={CH-1211},
             country={Switzerland}}
 
\affiliation[5]{organization={Physics Department, University of Pisa},
             addressline={Largo Pontecorvo 3},
             city={Pisa},
             postcode={56127},
             country={Italy}}


\begin{abstract}
The computation of the three-particle correlation function involving three hadrons started just recently after the first publications of ALICE measurements. Key elements to be considered are the correct description of the asymptotics, antisymmetrization issues and, in most cases, the treatment of the Coulomb interaction. In the case of the $ppp$ correlation function, a first analysis was done
where the hyperspherical adiabatic method was used to determine the $ppp$ wave function at different energies. Although the asymptotic behavior, antisymmetrization issues and the treatment of the Coulomb interaction were discussed in detail, the convergence properties of the adiabatic basis were studied at low energies around the formation of the correlation peak determined mainly by the $J^\pi=1/2^-$ and $3/2^-$ three-body states. Since many and very precise data have been taken or are planned to be measured at energies beyond the peak, we present an analysis of the convergence characteristics of the basis as the energy of the process increases. We show that in order to describe correctly the correlation tail it is necessary to consider three-body states up to $J^\pi=21/2^-$ whereas higher states can be considered as free. Once those states are incorporated solving the associate dynamical equations, the agreement with the experimental data is found to be excellent.
\end{abstract}

\begin{keyword}
Femtoscopic correlation functions \sep Three-body systems \sep Hyperspherical adiabatic expansion



\end{keyword}

\end{frontmatter}




\section{Introduction}

As a result of high-energy collisions at the Large Hadron Collider, particles are produced at very short relative distances, of the order of
a femtometer, that is, of the order of the range of the nuclear force. Thanks to the femtoscopy technique \cite{wie99,hei99,lis05,lfabb2021},
the correlation between the emitted particles can nowadays be measured. Interestingly, the correlation pattern is sensitive to the interaction between hadrons giving a new opportunity and an alternative way to study how hadrons interact among themselves at relative low energies~\cite{lfabb2021}. Moreover, in some cases traditional scattering experiments are not available, and therefore the correlation function emerges as an important mechanism to put in evidence characteristics of the two- or three-hadron interaction.
In particular, experimental information about the $pd$ and $ppp$ correlation functions is already available \cite{femtopd,femtoppp}. It is worth mentioning that scattering experiments in which three nucleons collide in the initial state are not yet possible in terrestrial laboratories. Accordingly the study of the $ppp$ correlation function will give the opportunity for a direct study of this process.

In a recent work \cite{kie24}, the difficulties of a theoretical description of the three-body $ppp$ correlation function were pointed out.  The continuum three-body
problem is always a multichannel problem, even if only central two-body potentials are involved. This fact complicates the correct normalization of
the scattering wave function since it requires an accurate calculation of the $S$-matrix, whose dimension is given by the often large number of
relevant coupled channels.  For the case of three charged particles, the extraction of the $S$-matrix presents the additional complication
coming from the fact that there is not a closed form describing the three-body Coulomb asymptotics.

In this work, we revisit the calculations made in Ref.~\cite{kie24}, emphasizing the role played by higher partial waves in the construction of the correlation function. At the two-body level,
it has been found that the main features of the $pp$ correlation function are captured by the lower partial waves~\cite{femtopp}. Actually, in Refs.~\cite{kie24,pdtheory}, it is shown that the inclusion of the nuclear potential only in the $s$-wave relative channel is sufficient to nicely reproduce the experimental data around the peak appearing at relatively low energy and, as the energy increases, other partial waves start playing a role.
In the $pd$ case a detailed analysis of the importance of the different partial waves in the construction of the correlation function was made in Ref.~\cite{pdtheory}. Here we will
show that for the $ppp$ case the same tendency appears, although the convergence in terms of partial waves is clearly slower. As the energy increases, the inclusion of a quite large number of partial waves is necessary in the description of the wave function to reach convergence in the correlation function. In the case of two outgoing particles the partial waves are related to the relative angular momentum which controls the centrifugal barrier. However in the three-body case the partial waves are related to the grand angular momentum, $K$, which controls the centrifugal barrier produced by the relative angular momentum between two of the particles and the relative angular momentum between the third particle and the center of mass of the first two. This makes the analysis of the convergence more complicated. To our knowledge this is the first time the impact of the short-range interaction is studied in the construction of the three-body scattering state as a function of the relative energy.

\section{Numerical method}

Three-body systems are often described by means of the Jacobi coordinates, $\bm{x}$ and $\bm{y}$, which are employed
to construct the so-called hyperspherical coordinates. These coordinates involve one radial coordinate, the hyperradius $\rho=\sqrt{x^2+y^2}$, and five hyperangles, $\Omega_\rho$, 
which contain the four angles describing the directions of $\bm{x}$ and $\bm{y}$ and $\alpha=\arctan x/y$ (for the precise
 definition of the Jacobi coordinates used in this work see Ref.~ \cite{kie24}).

Using these coordinates, for equal mass particles, the Koonin-Pratt formula \cite{Koonin,Pratt} can be generalized to three particles, and  the three-body correlation function 
becomes \cite{kie24}
\begin{equation} 
C_{123}(Q)=\int \rho^5 d\rho \,d\Omega_\rho \,S_{123}(\rho) |\Psi_s|^2\ ,
\label{c123}
\end{equation} 
where  $S_{123}(\rho)$ is the source function  which gives the emission probability of the three particles at a given value of $\rho$, 
and $|\Psi_s|^2$ is the square of the scattering wave function of the three particles. The three-body momentum, $Q$, introduced
in Eq.(\ref{c123}) is given by $Q^2=k_x^2+k_y^2$, where $\bm{k}_x$
and $\bm{k}_y$ are the conjugate momenta associated to the $\bm{x}$ and $\bm{y}$ Jacobi coordinates. Note that $Q$ differs
from the Lorentz invariant definition \cite{gra22}, usually denoted as $Q_3$. The relation between both quantities is in general
not simple, but for three identical particles, with the Jacobi coordinates used in Ref.~\cite{kie24}, this relation reduces to  $Q_3=\sqrt{6}\,Q$.

As shown in Ref.~\cite{kie24}, and considering a Gaussian source function for each proton, the source function for the $ppp$ system takes the form 
\begin{equation}
S_{123}(\rho)=\frac{1}{\pi^3 \rho_0^6}e^{-\rho^2/\rho_0^2}\ ,
\end{equation}
where the source size $\rho_0$ is given by $\rho_0=2R_M$, $R_M$ being the size of the proton source function.

In this work, the three-body scattering wave function, $\Psi_s$, is computed by means of the hyperspherical adiabatic expansion method, which is 
described in detail in Ref.\cite{nie01}. In Refs.~\cite{gar12,gar14}, this method is specifically employed to describe scattering wave functions.
Within this method, the wave function is expanded in terms of a basis set $\{\Phi_n^{J M}(\rho, \Omega_\rho)\}$, called the hyperpherical adiabatic basis which, for a given total angular momentum 
$J$ (with third component $M$), is
obtained as the set of eigenfunctions of the hyperangular part of the Schr\"{o}dinger 
equation \cite{nie01}.  
 
The basis functions $\Phi_n^{JM}(\rho,\Omega_\rho)$ are expanded as
 \begin{equation}
 \Phi_n^{JM}(\rho,\Omega_\rho)=\sum_{K, q} C_{K,q}^{J(n)}(\rho) \left[{\cal Y}_{\ell_x \ell_y}^{KL}(\Omega_\rho)\otimes \chi_S^{s_x \frac{1}{2}}\right]^{JM}\ ,
 \label{hhexp}
 \end{equation}
 where $q\equiv \left\{ \ell_x, \ell_y, L, s_x, S\right\}$. The quantum numbers  $\ell_x$ and $\ell_y$ are the orbital angular momenta associated with the $\bm{x}$ and
 $\bm{y}$ Jacobi coordinates, which couple to the total orbital angular momentum $L$. The spins of the two particles connected by the $\bm{x}$-coordinate
 couple to $s_x$, which in turn couples to the spin of the third particle, $\frac{1}{2}$, to the total spin $S$. Finally, $L$ and $S$ couple to the total angular momentum
 of the three-body system $J$.  In Eq.(\ref{hhexp}) the ${\cal Y}_{\ell_x \ell_y}^{KLM_L}(\Omega_\rho)$ functions are the usual hyperspherical harmonic (HH) functions, where 
 $K=2\nu+\ell_x+\ell_y$  ($\nu=0,1,2,\cdots$) is
 the grand-angular (or hyperangular) quantum number, and $ \chi_{S M_S}^{s_x \frac{1}{2}}$ is the spin part of the three-body wave function ($M_L$ and $M_S$ are the third components
 of $L$ and $S$, respectively).
Finally, within this model, as shown in Ref.~\cite{gar14}, the three-body scattering wave function takes the form
\begin{equation}
\Psi_s=\frac{(2\pi)^3}{(Q\rho)^{5/2}}\sum_{J M}\sum_n u_n^J(\rho,Q)\sum_{s_x S M_S} \langle \chi_{SM_S}^{s_x \frac{1}{2}} | \Phi_n^{JM}(Q,\Omega_Q)\rangle^*\ ,
\label{wfexp}
\end{equation}
with
\begin{equation}
u_n^J(\rho,Q)=\sum_{n'} u_{n}^{n'}(\rho,Q) \Phi_{n'}^{JM}(\rho,\Omega_\rho)\ .
\label{wfexp2}
\end{equation}
In Eqs.~(\ref{wfexp}) and~(\ref{wfexp2})
$n$ and $n'$ are the incoming and outgoing channels, and $\Omega_Q$ collects the hyperangles, analogous to $\Omega_\rho$, 
but in momentum space.

It is important to note here that, provided the two-body potentials do not diverge faster than $1/r^2$ at the origin, the sum over $K$ in 
 Eq.(\ref{hhexp}) is not present for $\rho=0$ \cite{nie01}. This means that for $\rho=0$ each basis term $n$  in Eq.(\ref{hhexp}) has a $K$-value associated, i.e., the one characterizing 
 the function $\Phi_n^{JM}$ at $\rho=0$.  Therefore, the sums over $n$ and $n'$ in Eqs.(\ref{wfexp}) and (\ref{wfexp2}) 
 (and the indices $n$ and $n'$ in the radial wave functions) can be replaced by $K$, and $K'$, understood
 as the grand-angular quantum numbers associated to $n$ and $n'$ at $\rho=0$.

The radial functions, $u_n^{n'}(\rho,Q)$ (or $u_K^{K'}(\rho,Q)$), are obtained as the solutions of a coupled set of second order differential equations, where the eigenvalues of the hyperangular 
part of the Schr\"{o}dinger equation enter as effective potentials  (see Ref.~\cite{nie01} for details). As shown in Ref.~\cite{kie24}, when only short-range 
potentials are involved, these wave functions have to be computed imposing the asymptotic behaviour
\begin{eqnarray}
\lefteqn{\hspace*{-3mm}
u_n^{n'}(\rho,Q)\equiv u_K^{K'}(\rho,Q) \stackrel{\rho \rightarrow \infty}{\longrightarrow}  } \nonumber \\ & &
i^{K'}\sqrt{Q\rho} \left( \delta_{KK'} J_{K'+2}(Q\rho)+ T_{KK'} {\cal O}_{K'+2} (Q\rho) \right)\ ,
\label{asym}
\end{eqnarray}
where $K$ and $K'$ are the grand-angular momentum values associated with the incoming and
outgoing channels $n$ and $n'$, respectively, ${\cal O}_{K+2}(Q\rho)=Y_{K+2}(Q\rho)+i J_{K+2}(Q\rho)$ is the
the outgoing asymptotic wave function,
$T_{KK'}$ is a $T$-matrix element, and $J_{K+2}(Q\rho),Y_{K+2}(Q\rho)$ are the regular and 
irregular Bessel functions.  

The asymptotic behavior in Eq.(\ref{asym}) guarantees that, in the case of no interaction, since  $T_{KK'}=0$, the expansion in Eq.(\ref{wfexp}) 
reduces to the partial wave expansion of the plane wave. In other words, in the free cases we have that  
$\Psi_s=e^{\bm{k}_x\cdot \bm{x}+\bm{k}_y \cdot \bm{y}} \sum_{S M_S s_x}\chi_{sM_S}^{s_x \frac{1}{2}}$,
and the correlation function in Eq.(\ref{c123}) is equal to 1 for all $Q$ values.

If the three-body system contains more than one charged particle, due to the presence of the Coulomb interaction, the asymptotic form 
(\ref{asym}) is no longer valid. In fact, there is no analytic expression analogous to Eq.(\ref{asym}) given in terms of Coulomb functions that would
describe the asymptotic behaviour of a continuum three-body wave function containing more than one charged particle. One possibility to
alleviate this problem, known to work well when computing correlation functions, is to screen the Coulomb potential, such that all the interactions
are in practice short-range \cite{gar24}. However, in the case of three identical charged particles, an efficient alternative is to average the
Coulomb interaction over the hyperangles in such a way that the full Coulomb potential is reduced to the $\rho$-dependent form
$V_\mathrm{Coul}(\rho)=\frac{16}{\sqrt{2}\pi}\frac{e^2}{\rho}$ \cite{kie24}. Doing like this the asymptotic form (\ref{asym}) is still valid simply
replacing the regular and irregular Bessel functions, $J_{K+2}(z)$ and $Y_{K+2}(z)$, by $\sqrt{2/(\pi z)}F_{K+\frac{3}{2}}(\eta,z)$ and 
$\sqrt{2/(\pi z)}G_{K+\frac{3}{2}}(\eta,z)$, where $F$ and $G$ are the regular and irregular Coulomb functions, and the Sommerfeld
parameter is given by $\eta=16 m e^2/(\pi \hbar^2 Q)$.

\section{Accuracy requirements}

In the present calculation, it is crucial to ensure that the scattering wave function is obtained with sufficient accuracy. To this aim, 
the scattering wave function given in Eq.~(\ref{wfexp}) is divided in two pieces, $\Psi_s=\Psi_s^\mathrm{comp}+\Psi_s^\mathrm{free}$. 
The first part, $\Psi_s^\mathrm{comp}$, is given 
by Eq.(\ref{wfexp}), but containing only the computed radial wave functions, 
$u_n^{n'}(\rho,Q)$, for all the adiabatic channels, $n$ and $n'$, associated to grand-angular quantum 
numbers $K\leq K_0$. The quantum number $K_0$ thus limits the number of radial wave functions computed numerically. 

\begin{table*}[t!]
	\setlength\tabcolsep{.25\tabcolsep}
\begin{tabular}{c|ccccccccccc|cccccccccc}
& $\frac{1}{2}^-$ & $\frac{3}{2}^-$ & $\frac{5}{2}^-$ & $\frac{7}{2}^-$ & $\frac{9}{2}^-$ & $\frac{11}{2}^-$ & $\frac{13}{2}^-$ & $\frac{15}{2}^-$ & $\frac{17}{2}^-$&$\frac{19}{2}^-$  &$\frac{21}{2}^-$
& $\frac{1}{2}^+$ & $\frac{3}{2}^+$ & $\frac{5}{2}^+$ & $\frac{7}{2}^+$  & $\frac{9}{2}^+$ & $\frac{11}{2}^+$ & $\frac{13}{2}^+$ & $\frac{15}{2}^+$  &  $\frac{17}{2}^+$& $\frac{19}{2}^+$  \\ \hline
$n \rightarrow K$=1 & 1  &  1 &     &        &       &       &     &     &     &     &    &  &  &   &  &  &  &  &  & & \\
$n \rightarrow K$=2 &     &     &      &        &       &       &     &     &     &     &    & 2 & 2 & 2  &  &  &  &  &  & &\\
$n \rightarrow K$=3 & 2 & 4  & 4   & 2    & 1    &       &     &    &      &     &    &  &  &   &  &  &  &  &  & &  \\
$n \rightarrow K$=4 &    &     &       &        &       &       &     &    &      &     &    & 3 & 5 & 5  &5  & 3 &  &  &  & & \\
$n \rightarrow K$=5 & 4 & 6 & 8    & 8    & 6    &  4  & 1  &    &      &     &    &  &  &   &  &  &  &  &  & &\\
$n \rightarrow K$=6 &    &     &       &        &       &       &     &    &      &     &    & 4 & 8 & 10  & 10 & 10 & 8 & 4 & 1 & &\\
$n \rightarrow K$=7 & 5 & 9 & 11 & 13  & 13 & 11 & 9  & 5 & 1  &     &    &  &  &   &  &  &  &  &  & & \\ 
$n \rightarrow K$=8 &   &   &    &     &    &    &    &   &    &     &    & 6 & 10  & 14  & 16 & 16  & 16  & 14 & 10 & 6 & 1\\ 
$n \rightarrow K$=9 & 6 & 12 & 16 & 18  & 20 & 20 & 18  & 16 & 12  &  6   &  2  &  &  &   &  &  &  &  &  & & \\ \hline
Total $K_0$=1 & 1  &  1 &     &       &      &       &    &     &    &    &    &     &  &   &  &   &   &  &  & & \\
Total $K_0$=2 & 1  &  1 &     &       &      &       &    &     &    &    &    &  2& 2 &  2 &   &  &  &  &  & &\\
Total $K_0$=3 & 3  &  5 & 4   &  2    &  1   &       &       &    &     &     &    &  2& 2 &  2 &   &  &  &  & & & \\
Total $K_0$=4 & 3  &  5 & 4   &  2    &  1   &       &      &     &     &     &    &  5& 7 &  7 &  5 & 3 &  &  & & & \\
Total $K_0$=5 & 7  & 11 & 12   &  10    &  7   &  4     & 1  &     &     &    &    &  5& 7 &  7 &  5 & 3 &  &  & & & \\
Total $K_0$=6 & 7  & 11 & 12   &  10    &  7   &  4     & 1  &     &     &    &    &  9& 15 &  17 &  15 & 13 & 8 & 4 & 1 & &\\
Total $K_0$=7 & 12  &  20 &   23  &     23   &     20  &     15  &   10  & 5    & 1   &      &   &  9& 15 &  17 &15  & 13 & 8 & 4 & 1 & &\\
Total $K_0$=8 & 12  &  20 &   23  &     23   &     20  &     15  &   10  & 5    & 1   &      &   &  15& 25 &  31 &31  & 29 & 24 & 18 & 11 & 6 & 1\\
Total $K_0$=9 & 18  &  32 &   39  &     41   &     40  &     35  &   28  & 21    & 13   &  6    &   2 &  15& 25 &  31 &31  & 29 & 24 & 18 & 11 & 6 & 1\\ \hline
$N_\mathrm{min}$($K_0$=1) & 198  &  198 &     &       &      &       &    &     &    &    &    &     &  &   &  &   &   &  &  & & \\
$N_\mathrm{min}$($K_0$=2) &     &     &     &       &      &       &    &     &    &    &    &  326& 455 &  325 &   &  &  &  &  & &\\
$N_\mathrm{min}$($K_0$=3) & 458  &  718 & 716   &  455    &  130   &       &       &    &     &     &    &   &   &    &   &  &  &  & & & \\
$N_\mathrm{min}$($K_0$=4) &     &     &       &         &       &       &      &     &     &     &    &  582& 967 &  1093 &  961 & 576 &  &  & & & \\
$N_\mathrm{min}$($K_0$=5) & 714  & 1230 & 1484   &  1479    &  1218   &  704     & 192  &     &     &    &    &   &    &    &    &   &  &  & & & \\
$N_\mathrm{min}$($K_0$=6) &     &       &         &           &        &         &     &     &     &    &    &  834& 1471 &  1849 &  1969 & 1836 & 1451 & 819 & 189 & &\\
$N_\mathrm{min}$($K_0$=7) & 966  &  1734 &   2240  &     2487   &     2478  &     2216  &   1704  & 945    & 252   &      &   &   &     &      &     &    &   &   &   & &\\
$N_\mathrm{min}$($K_0$=8) &        &        &          &            &             &            &         &        &      &      &   &  1082& 1967 &  2593 &2961  & 3076 & 2939 & 2555 & 1925 & 1054 & 248\\
$N_\mathrm{min}$($K_0$=9) & 1214  &  2230 &   2984  &     3479   &     3718  &     3704  &   3440 & 2929    & 2174   &  1178    &   310 &    &     &      &     &     &     &     &     &     &    \\ \hline
\end{tabular}
\caption{The upper part of the table shows the number of adiabatic terms, $n$, associated to given values of the grand-angular quantum number $K$, for three identical fermions with spin $\frac{1}{2}$ and for each $J^\pi$-state. The central part lists, for different values of $K_0$, the total number of channels for each $J^\pi$-state, which gives the dimension of the $(u_n^{n'})$ matrix in Eq.(\ref{wfexp2}). The lower part of the table shows $N_\mathrm{min}(K_0)$ which is, for each $K_0$, the
minimum number of terms to be included in Fabbiettithe expansion given in Eq.~(\ref{hhexp}) of the angular eigenfunctions, $\Phi_n^{JM}$, when $K_\mathrm{max}=130$. }
\label{tab1}
\end{table*}

In the upper part of Table~\ref{tab1}, we give, for the case of three identical spin-$\frac{1}{2}$ fermions, the number of adiabatic channels associated with a given $K$-value, up to $K$=9, for the different $J^\pi$ states. As it is obvious from the definition of $K$ ($K=2\nu+\ell_x+\ell_y$, with $\nu=0,1,2,\cdots$),  
negative and positive parity states are associated with odd and even values of $K$, respectively. The central part of the table gives the total number of channels for each $J^\pi$ state and different values of $K_0$. This number is, therefore, the dimension of the $(u_n^{n'})$ matrix of radial wave functions 
in Eq.(\ref{wfexp2}) to be computed for each $J^\pi$ and $K_0$. As we can see from the table, this
number is obviously given by the sum of the corresponding adiabatic channels
up to the chosen $K_0$ value.

The remaining part of the scattering wave
function, $\Psi_s^\mathrm{free}$, is included assuming free radial wave functions, i.e. 
$u_n^{n'}(\rho,Q)=i^{K'} \sqrt{Q\rho} J_{K'+2}(Q\rho) \delta_{K K'}$, which leads to
\begin{eqnarray}
\Psi_s^\mathrm{free}&=&\frac{(2\pi)^3}{(Q\rho)^2}\sum_{J M}\sum_{K>K_0}   i^K  J_{K+2}(Q\rho) \Phi_{n'}^{J}(\rho,\Omega_\rho)   \nonumber \\ & &
\times \sum_{s_x S M_S} \langle \chi_{SM_S}^{s_x \frac{1}{2}} | \Phi_n^{J}(Q,\Omega_Q)\rangle^*\ .
\label{freep}
\end{eqnarray}
Here, as described above, $J_{K+2}(z)$ has to be replaced by $\sqrt{2/(\pi z)}F_{K+\frac{3}{2}}(\eta,z)$, when dealing with three identical charged particles 
and the Coulomb potential is averaged over the hyperangles.

To evaluate Eq.(\ref{freep}), although it is still necessary to cut the sum over $K$ at some large value, since the expression is fully analytic, this maximum value can be easily taken sufficiently high (typically $K$ values up to 30 or 40 are more than enough for $Q$-values up to about 1 GeV/$c$). On top of that, it is also necessary to know how many times each $K$-value enters in the sum. In general,
this is given by the simple formula $N(K)=(K+1)(K+2)^2(K+3)/12$, the number of HH functions having the same $K$ value. It is however crucial to note that, whereas the computed wave functions included 
in $\Psi_s^\mathrm{comp}$ are obtained with the correct symmetry requirements (full antisymmetrization in the case of three identical fermions), the expression in Eq.(\ref{freep}) does not in principle include these requirements. This means that when evaluating the number of times that each $K$-value enters in Eq.(\ref{freep}), one has to know how many of the $N(K)$ states given above actually have the correct symmetry.
For instance, whereas $N$($K$=1)=6, $N$($K$=2)=20, or $N$($K$=3)=50, from the upper part of  Table~\ref{tab1} we can see that for three identical spin-$\frac{1}{2}$ fermions
the $K$=1, $K$=2, and $K$=3 values appear only 2, 6, and 13 times, respectively, as a result of imposing the correct symmetry. The number of $K$-states
with the desired symmetry can be obtained analytically as described in Ref.~\cite{kie24}. 

Together with the $K_0$-value that determines the number of radial wave functions computed numerically in the expansion in Eq.(\ref{wfexp}), it is also necessary
to make sure that the basis functions, $ \Phi_n^{JM}$, are accurately computed, i.e., that the expansion in Eq.(\ref{hhexp}) has converged as well. This is done by
including in the sum in Eq.(\ref{hhexp}) all the terms with $K\leq K_\mathrm{max}$, as well as, at least, all the components, $q\equiv\{ \ell_x, \ell_y,L,s_x,S\}$,
with $\ell_x$ and $\ell_y$ contributing to the chosen $K_0$-value, that is, fulfilling the condition  $\ell_x+\ell_y \leq K_0$. This implies
that the larger the $K_0$ value, the larger the minimum number of terms required in the expansion of Eq.~(\ref{hhexp}). In the lower part of Table~\ref{tab1}, the value of $N_\mathrm{min}(K_0)$ denotes, for each $J^\pi$ state, the minimum number of terms in the expansion of Eq.~(\ref{hhexp}) for 
a given $K_0$ and $K_\mathrm{max}=130$,
which is  a value such that, as shown in Ref.~\cite{kie24}, the $ppp$ correlation function has converged.

The fact that the convergence has to be reached at two different levels, in Eq.(\ref{hhexp}) and in Eq.(\ref{wfexp}), is in fact the main advantage of using the hyperspherical adiabatic expansion method. From the numerical point of view, the value of $K_\mathrm{max}$ used to cut the sum in Eq.~(\ref{hhexp}) can be very large. It is not difficult to deal with expansions of the $\Phi_n^{JM}$ basis functions containing several thousands of terms. Furthermore, it should be remarked that the $\Phi_n^{JM}$ functions do not 
depend on the three-body momentum, $Q$, and they have to be computed only once for each $J^\pi$ state. On the contrary, as we will see, convergence in the results will be obtained with relatively modest values of $K_0$, which implies that the number of radial wave functions, $u_n^{n'}(\rho,Q)$, to be computed and normalized according to Eq.(\ref{asym}) is still manageable. As an example, as seen in Table~\ref{tab1}, for three identical spin-$\frac{1}{2}$ fermions, and using $K_0=7$, the worst scenario corresponds to the calculation of the $\frac{5}{2}^-$ and $\frac{7}{2}^-$ states, where up to 23 adiabatic terms have to be included, being therefore necessary to solve a coupled set of 23 differential 
equations that permit to obtain the required  $23^2=529$ radial wave functions, $u_n^{n'}(\rho,Q)$. For these two states, the minimum number of terms in the 
expansion of Eq.~(\ref{hhexp}) is rather large, about 2500, but still acceptable.

All this contrasts with the direct expansion of the scattering function in terms of the hyperspherical harmonics. Convergence of the results very likely 
requires the inclusion of several thousands of hyperspherical harmonics in the expansion. Therefore, in this case, it would be necessary to solve a system of several thousands of coupled differential equations and from them extract millions of radial wave functions with the proper normalization. Very soon, this procedure becomes completely unfeasible. 

In any case, even if the method described here allows us to compute three-body correlation functions much more efficiently than with the
hyperspherical harmonic expansion, we can see from Table~\ref{tab1} that the required
numerical effort increases rather fast. The change from $K_0$=7 to $K_0$=9 significantly increases the number of radial functions
to be computed and the number of terms in the expansion in Eq.(\ref{hhexp}) (about 1600 radial functions and 3500 terms for the 
$\frac{5}{2}^-$, $\frac{7}{2}^-$, $\frac{9}{2}^-$ states). Although the expansion in Eq.(\ref{hhexp}) has to be computed only once, the
radial functions have to be obtained for each value of the three-body momentum $Q$, and this dramatically increases the computation
time. In this work, we have limited our calculations to $K_0$ values up to 7.

\section{The $ppp$ correlation function}

In Ref.~\cite{kie24}, the $ppp$ correlation function was computed following the procedure described above. As shown in that work, a $K_\mathrm{max}$ value of about 130 is needed to get convergence in the correlation function, mainly for low values of the three-body momentum. At the two-body level,
it was seen that the experimental $pp$ correlation function, dominated by the low-energy peak, was nicely reproduced including the nuclear force only for $s$-waves, and taking the remaining 
partial waves as free. For this reason, in Ref.~\cite{kie24}, since the effort was focused on the description of the peak, the value $K_0=2$ was found to be sufficient to obtain an accurate enough $ppp$ correlation function in the low-energy region.
According to Table~\ref{tab1}, $K_0$=2 implies that the $\frac{1}{2}^-$,   $\frac{3}{2}^-$,  $\frac{1}{2}^+$,  $\frac{3}{2}^+$, and  $\frac{5}{2}^+$
states are computed into $\Psi_s^\mathrm{comp}$, whereas the remaining waves are included as free waves in $\Psi_s^\mathrm{free}$ 
($\Psi_s=\Psi_s^\mathrm{comp}+\Psi_s^\mathrm{free}$).  

In the calculation, the Argonne $v_{18}$ proton-proton potential was used \cite{AV18}.
The inclusion of a three-body force, not considered in the present study, was analysed in Ref.\cite{kie24} observing a contribution to the $ppp$ correlation function of the order of 1\% or less. Furthermore,
the Coulomb
interaction was averaged over the hyperangles, which, as mentioned above, transforms the Coulomb potential into the $\rho$-dependent interaction
$V_\mathrm{Coul}(\rho)=\frac{16}{\sqrt{2}\pi}\frac{e^2}{\rho}$.

When comparing to the experimental data \cite{femtoppp}, we obtained what is shown in Fig.12 of Ref.~\cite{kie24}.
The computed results were given
for different values of the source size $\rho_0$, and the possibility of detecting secondary protons coming from the decay of hyperons like the
$\Lambda$-particle was taken into account. As it can be seen from that figure, the computed correlation functions are systematically below the experimental data, and,
even more, they reach the asymptotic value of 1 from below, whereas the data do it from above.

In Ref.~\cite{kie24}, that discrepancy was attributed either to the modelling of the three-body source function, which is currently derived from two-body femtoscopy measurements, or to the need of a more detailed comparison between the data and the theory in the region where the data are normalised to the unity at $Q_3 > 1$ GeV/$c$.
\begin{figure}[t]
\centering
        \includegraphics[scale=0.35]{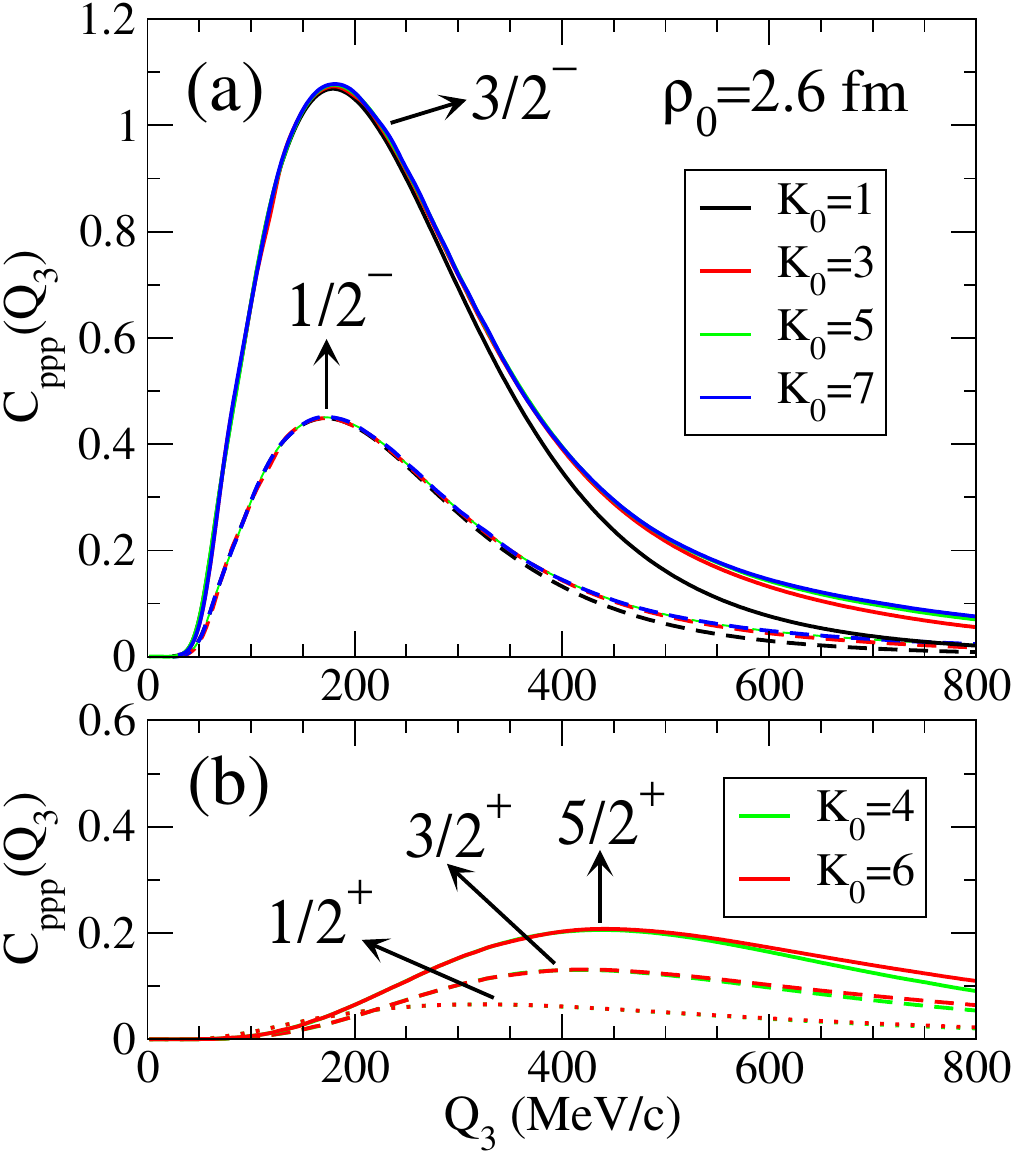}
        \caption{(a) Contribution to the correlation function from the $\frac{1}{2}^-$ (dashed) and $\frac{3}{2}^-$ (solid) states for $K_0=1, 3, 5$ and 7.
        (b) The same for the    $\frac{1}{2}^+$ (dotted), $\frac{3}{2}^+$ (dashed), and $\frac{5}{2}^+$ (solid) states for $K_0=4$ and 6. The calculations have
        been made using a source size of $\rho_0$=2.6 fm. }
        \label{fig:conv}
\end{figure}
A deep analysis of the pattern of convergence has shown that the value  $K_0=2$, the maximum grand angular quantum number considered in the description of the interacting system,
is actually too small, giving only a reasonable description around the peak. In Fig.~\ref{fig:conv} we show for the $\frac{1}{2}^-$ and $\frac{3}{2}^-$ states (panel (a)), and for the 
$\frac{1}{2}^+$, $\frac{3}{2}^+$, and $\frac{5}{2}^+$ states (panel (b)), how their contribution to the correlation function changes when the value
 of $K_0$ is increased. To keep panel (b) clean, we have omitted the curves corresponding to $K_0=2$. The calculations
 correspond to a source size $\rho_0$=2.6 fm.
\begin{figure}[t]
\centering
        \includegraphics[scale=0.35]{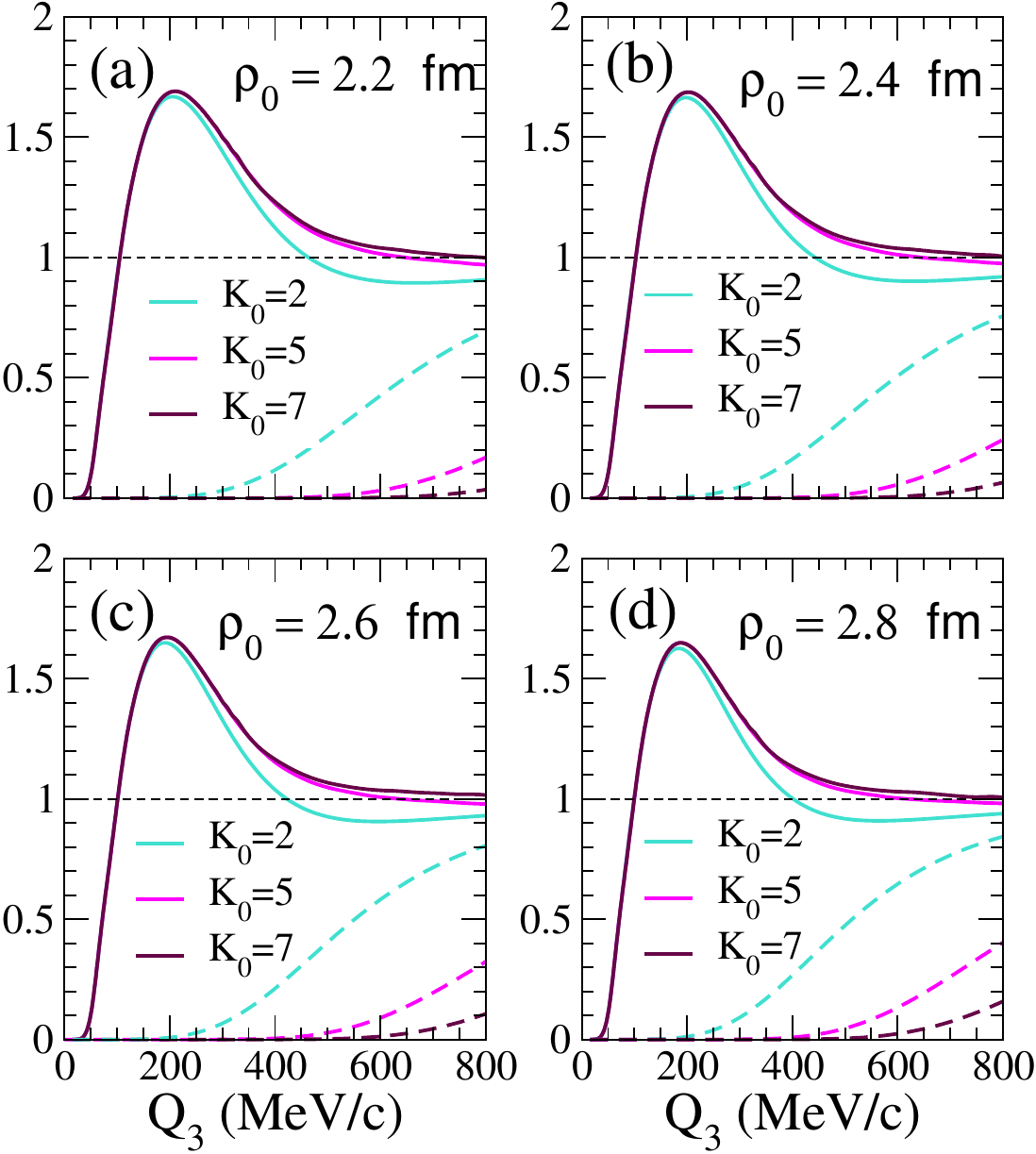}
        \caption{Total $ppp$ correlation function (solid curves) for (a) $\rho_0$=2.2 fm,   (b) $\rho_0$=2.4 fm, (c) $\rho_0$=2.6 fm,  and (d) $\rho_0$=2.8 fm, 
        when $K_0=2,5$ and 7. The dashed curves show the contribution to the correlation function from the free
        part of the wave function, $\Psi_s^\mathrm{free}$, for each of the cases.}
        \label{fig:conv-full}
\end{figure}

As we can see by inspection of the figure, the variation produced in the $\frac{1}{2}^-$ and $\frac{3}{2}^-$ contributions when going from $K_0=1$ (black curves) to $K_0=3$ (red curves)
is clearly visible for $Q_3$ values above about 350 MeV$/c$. Even the increase of $K_0$ up to 5 (green curves) still produces a small increase on the contribution
to the correlation function, particularly in the $\frac{3}{2}^-$ case, beyond 600 MeV$/c$. When going from $K_0$=5 to $K_0$=7, the increase in the correlation
function is already tiny. For the states where the calculation with $K_0$=9 can be easily performed, namely the $\frac{17}{2}^-$, $\frac{19}{2}^-$, and $\frac{21}{2}^-$ states, we have observed that the variation in their contribution to the correlation function when compared to their contribution as free waves is negligible, of the order of $10^{-4}$ for large values 
of $Q_3$, where the peak of their contribution is located.
By inspection of the lower part of the figure, 
considering the modest change produced in the  $\frac{3}{2}^+$ and $\frac{5}{2}^+$ contributions when going from $K_0=4$ to $K_0=6$ (for the
$\frac{1}{2}^+$ case the change is actually not visible), we can foresee that increasing $K_0$ up to 8 would give rise to a negligible contribution. 
Therefore, performing the calculations with $K_0$=8 is already not worth it, considering the big numerical effort required to compute some of the states. 
For instance, as shown in Table~\ref{tab1}, for $K_0$=8 the  $\frac{5}{2}^+$ and $\frac{7}{2}^+$ states contain up to 31 relevant adiabatic channels, which makes the calculations of these states, including the nuclear potential, very heavy. 
As before, for the cases where the calculation with $K_0$=8 can be made ($\frac{15}{2}^+$, $\frac{17}{2}^+$, and $\frac{19}{2}^+$ states) the variation in their contribution to the correlation function, although bigger than in the $K_0=9$ case, is still very small, not larger than $10^{-3}$. 

We have then taken $K_0$=7 as the maximum value used in our calculations. In Fig.~\ref{fig:conv-full}, we show the total $ppp$ correlation function
for (a) $\rho_0$=2.2 fm, (b) $\rho_0$=2.4 fm, (c) $\rho_0$=2.6 fm, and (d) $\rho_0$=2.8 fm. We give the results for $K_0$=2, $K_0$=5, and $K_0$=7.
For each of these $K_0$ values, the dashed curves show the contribution to the correlation function from $\Psi_s^\mathrm{free}$, Eq.(\ref{freep}), that is, 
from the partial waves that are included as free, without the inclusion of the nuclear potential.

\begin{figure}[t!]
\centering
         \includegraphics[scale=0.35]{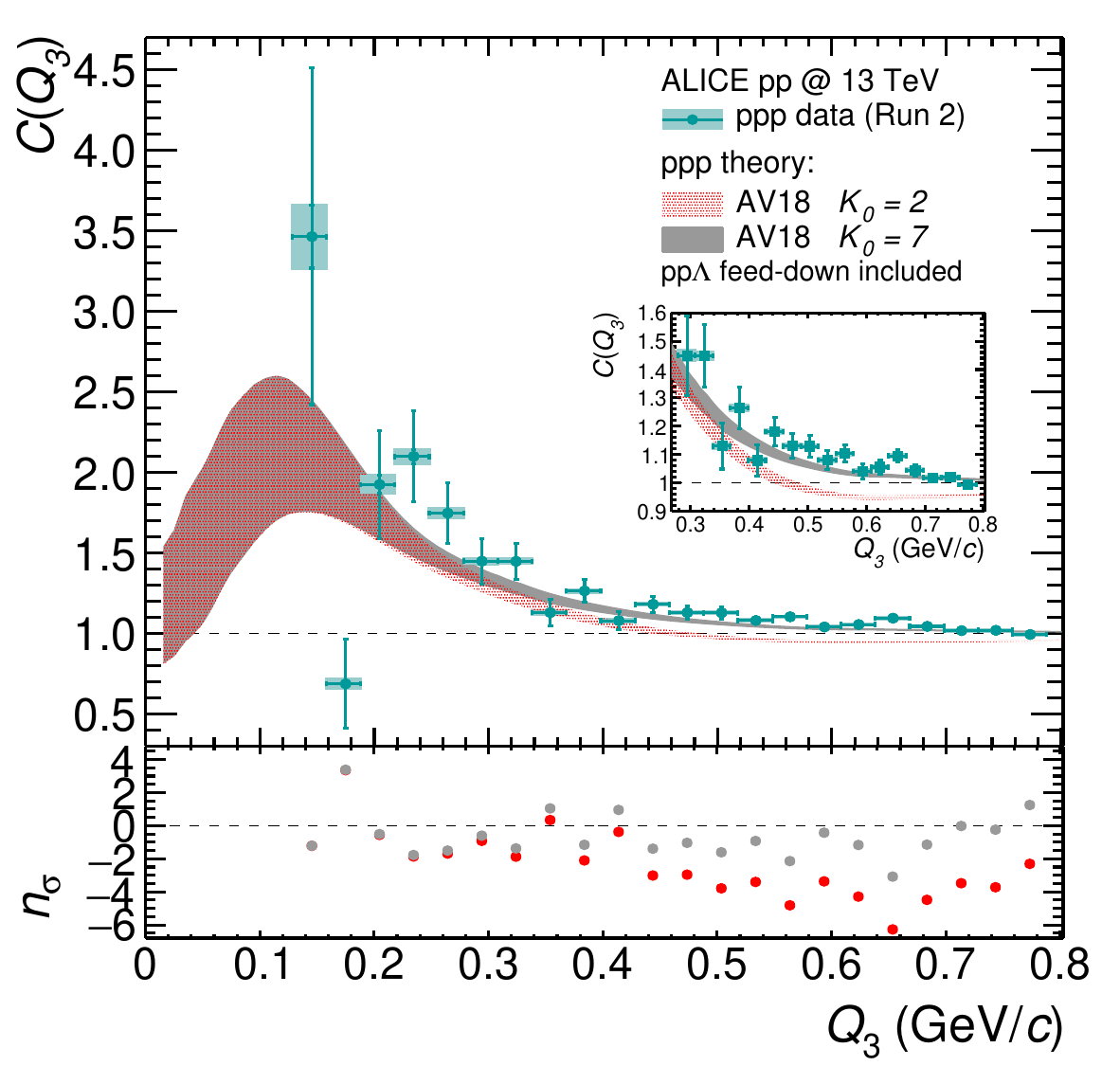}
        \caption{The comparison of the $ppp$ correlation function measured by the ALICE Collaboration~\cite{femtoppp} (cyan full squares) and the calculated correlation functions $K_0$=2 and $K_0$=7. The band of the theory curves includes the uncertainty of the pp$\Lambda$ feed-down, propagated from the experimental pp$\Lambda$ correlation function in ~\cite{femtoppp}, as well as the uncertainty on the source radius $\rho_0 = (2.5 \pm 0.1)$ fm. The bottom panel shows the deviation of the theory curves from the data. The latter is evaluated in number of standard deviations ($n_\sigma$), considering the uncertainties of the data (statistical and systematics) and the band of the theory.  }
        \label{fig:final-comp}
\end{figure}

As seen by inspection of the figure, the jump from $K_0$=2 to $K_0$=5 gives rise to a significant change in the correlation function for $Q_3$ values larger than about
200 MeV/$c$, in such a way that already for $K_0$=5, the function approaches 1 from above, as the experimental data do. We can however 
see that an additional increase of $K_0$ up to 7 still produces a small correction of the correlation function for large $Q_3$ values, making clear that the 
computed function has converged within the $Q_3$ range shown in the figure, therefore confirming that the computed correlation functions
agree with the experimental observation of approaching 1 from above for large values of $Q_3$. 

We finally show in Fig.~\ref{fig:final-comp} how the results obtained using $K_0$=2 and $K_0$=7 compare to the experimental data \cite{femtoppp} once the computed curves are
corrected to include the possibility of the experimental detection of protons coming from $\Lambda$ hyperon decay (feed-down), as explained in Ref.~\cite{kie24}. The feed-down contribution from $pp\Lambda$ has been evaluated by using the experimental $pp\Lambda$ correlation function measured in Ref.~\cite{femtoppp}. The band of the theory curves in Fig.~\ref{fig:final-comp} includes both the experimental uncertainties of the $pp\Lambda$ correlation function and the uncertainty on the source radius, the former giving the dominant contribution. The proton source radius, $R_M = (1.24 \pm 0.04)$ fm, has been evaluated from the resonance source model in Ref.~\cite{ALICEsource,ALICEsource2}, by using as input the transverse mass of the $pp$ pairs in the $ppp$ triplets given in Ref.~\cite{hepdata.134041}. The obtained value for $R_M$ corresponds to $\rho_0 = (2.5 \pm 0.1)$ fm. We can see that the agreement between theory and
experiment for $Q_3$ values larger than 0.3 GeV/$c$ is now excellent and much better than in Ref.~\cite{kie24}, where we limited the interaction to act up to $K_0=2$. Futhermore, the average deviation of the result with $K_0 = 7$ from the data, shown in the bottom panel of Fig.~\ref{fig:final-comp}, is found to be within 2$\sigma$.

\section{Conclusions}

In this work we have performed a study of the convergence of the $ppp$ correlation function in terms of three-body partial waves. To our knowledge studies of this type in nuclear physics do not exist since $ppp\rightarrow ppp$ scattering experiments are not feasible in terrestrial laboratories. In the case of the correlation function the three outgoing protons can be detected whereas the initial state is mimicked by the source function. Therefore, in order to describe the $ppp$ correlation function the scattering wave function of three protons has to be computed. Dividing the wave function in two parts, the interacting part and the free part, and considering the energies at which the correlations are measured, we have concluded that grand angular quantum numbers up to $K$=7 are required to describe the interacting part of the wave function. Above $K$=7 the wave function can be considered as free from the strong interaction. 

We have also shown in detail the technical difficulties to face when a large number of asymptotic states have to be considered. After their inclusion, the correlation function is in excellent agreement with the ALICE data. Since measurements of correlation functions involving three particles started to appear, the present analysis will be useful in their computation, as, for example, for the $pp\Lambda$ correlation function~\cite{garrido2024}.

\section*{Acknowledgements}
This work has been partially supported by: Grant PID2022-136992NB-I00 funded by MCIN/AEI/10.13039/501100011033.

\end{document}